**An observational correlation between stellar brightness variations and surface gravity**

Fabienne A. Bastien[1], Keivan G. Stassun[1,2], Gibor Basri[3], Joshua Pepper[1,4]

[1]Department of Physics & Astronomy, Vanderbilt University, 1807 Station B, Nashville, TN 37235

[2]Physics Department, Fisk University, 1000 17th Ave. N, Nashville, TN 37208

[3]Astronomy Department, University of California, Hearst Field Annex, Berkeley, CA 94720

[4]Physics Department, Lehigh University, 27 Memorial Dr. W, Bethlehem, PA 18015

**Surface gravity is one of a star's basic properties, but it is difficult to measure accurately, with typical uncertainties of 25-50 per cent if measured spectroscopically[1,2] and 90-150 per cent photometrically[3]. Asteroseismology measures gravity with an uncertainty of about two per cent but is restricted to relatively small samples of bright stars, most of which are giants[4,5,6]. The availability of high-precision measurements of brightness variations for >150,000 stars[7,8] provides an opportunity to investigate whether the variations can be used to determine surface gravities. The Fourier power of granulation on a star's surface correlates physically with surface gravity[9,10]; if brightness variations on timescales of hours arise from granulation[11], then such variations should correlate with surface gravity. Here we report an analysis of archival data that reveals an observational correlation between surface gravity and the root-mean-square brightness variations on timescales of less than eight hours for stars with temperatures of 4500-6750 K, log of surface gravities of 2.5-4.5 (cgs units), and having overall brightness variations <3 parts per thousand. A straightforward**

**observation of optical brightness variations therefore allows a determination of the surface gravity with a precision of <25 per cent for inactive Sun-like stars at main-sequence to giant stages of evolution.**

Brightness variations of Sun-like stars are driven by many factors, including granulation[12], oscillations[11], rotation and magnetic activity[13]. As they evolve from high surface gravity ($g$) dwarfs to low-$g$ giants, their convective zones deepen, they rotate more slowly, their magnetic activity diminishes, and their oscillation and granulation timescales increase, all of which will change the nature of the brightness variations. It has been previously demonstrated that the power in granulation (as traced by the Fourier spectrum of the brightness variations) is inversely proportional to $\nu_{max}$, the peak frequency of Sun-like acoustic oscillations[9,12]. Given that $\nu_{max}$ is itself proportional to $g$[11], it follows that $g$ should manifest in brightness variations on timescales that trace granulation. Although physically we expect this, it is not immediately apparent that brightness variations can be used as an effective determinant of $g$ because other phenomena not directly related to $g$—most importantly spots, plage and other sources of brightness variations driven by the star's magnetic activity—probably dominate the observed brightness variations. It is therefore necessary to filter out the brightness variations arising from these phenomena, which occur on timescales of hours to days, while preserving the brightness variations related to granulation and $g$ on timescales of minutes to hours.

Using long cadence (30 minute) light curves from Quarter 9 of NASA's Kepler Mission[14], and representing them using the Filtergraph data visualization tool[15], we observe clear patterns in the evolutionary properties of stars encoded in three simple measures of their brightness variations[8] (Fig. 1): Range ($R_{var}$), number of zero crossings ($X_0$), and root-mean-square on time scales shorter than 8 hours (to which we refer hereafter as "8-hr flicker", or $F_8$). Relating these measures to $g$ determined asteroseismically for a sample of Kepler stars[4], we find distinctive features that highlight the way stars evolve in this three-dimensional space, making up an evolutionary diagram of photometric variability. Within this diagram we find a vertical cloud of points, largely made up of high-$g$ dwarfs, that show large $R_{var}$, small $X_0$, and low $F_8$ values. We observe a tight sequence of stars—a "flicker floor" sequence that defines a prominently protruding lower envelope in $R_{var}$—spanning gravities from dwarfs to giants. Sun-like stars of all evolutionary states evidently move onto this sequence only when they have a large $X_0$, which in turn implies low stellar activity.

We find that $g$ is uniquely encoded in $F_8$, yielding a tight correlation between the two (Fig. 2). Moreover, using 11 years of SOHO Virgo[16,17] light curves of the Sun and sampling them at the same cadence as the Kepler long-cadence light curves, we find that the Sun's (constant) $g$ is also measurable using $F_8$, which remains invariant throughout the 11-year solar activity cycle even while the Sun's $R_{var}$ and $X_0$ change significantly from the spot-dominated solar maximum to the nearly spotless solar minimum. From the Sun's behavior we infer that a large portion of the Kepler stars' vertical scatter within the vertical cloud at the left of the diagram may be driven by solar-type cyclic activity variations. Most importantly, the Sun's true $g$ fits our empirical relation, and the $g$ value of any Sun-like Kepler star from dwarf to giant may be inferred from this relation with an accuracy of 0.06-0.10 dex (Supplementary Information).

Asteroseismic analyses derive $g$ from the properties of stellar acoustic oscillations[4,18,19,20]. Given that near-surface convection drives both these oscillations and granulation, and given the brightness variability time scales to which $F_8$ is sensitive, we suggest that a combination of different types of granulation (with typical solar time scales from ~30 minutes to ~30 hours[21]) drives the manifestation of $g$ in this metric. The precise time scales of these phenomena in solar-type stars depend strongly on the stellar evolutionary state and hence also on $g$[5,9,10,22]. Acoustic oscillations, whose amplitudes are sensitive to $g$[5], may provide an increasingly important contribution to $F_8$ as stars evolve into subgiants and giants and the amplitudes and time scales of these oscillations increase[5,9,10]. At some point, the pressure-mode and granulation time scales cross[9], which may lead to a breakdown of our $F_8$-$g$ relation at very low values of $g$.

By using $F_8$ to measure $g$, we can construct a photometric variability evolutionary diagram for most stars observed by Kepler, even for stars well beyond the reach of asteroseismic and spectroscopic analysis (Fig. 3). By coding this diagram according to the stellar temperature and rotation period, we may trace the physical evolution of Sun-like stars as follows: stars begin as main-sequence dwarfs with large photometric $R_{var}$ values and small $X_0$ values, presumably driven by simple rotational modulation of spots at relatively short rotation periods. As the stars spin down to longer rotation periods, their brightness variations first become steadily "quieter" (systematically lower $R_{var}$) but then become suddenly and substantially more complex (larger $X_0$) as they reach the flicker floor. Some stars reach the floor only after beginning their evolution as low-$g$ subgiants, having moved to the right (higher $F_8$) as their effective temperatures begin rapidly dropping. Other stars join the sequence while still dwarfs; these are easily identified in our diagram by the drastically increased $X_0$ at very low $F_8$. Evidently some dwarf stars become magnetically quiet while still firmly on the main sequence, whereas others do not reach the floor until they begin to swell considerably. We note that the Sun seems to approach the flicker floor at solar minimum; its $R_{var}$ value becomes quite low and its $X_0$ value strongly increases (Fig. 1).

A star's main-sequence mass and initial spin probably determine where along the flicker floor sequence it ultimately arrives, because the slope of a star's trajectory in our diagram is essentially determined by the ratio of its spin-down time scale (downward motion) and structural evolutionary time scale (rightward motion). Regardless, once on the floor all stars evolve along this sequence and stay on it as they move up to the red giant branch, their effective temperatures steadily dropping as their surfaces rapidly expand. Despite their very slow rotation as subgiants and giants on the flicker floor sequence, their photometric $R_{var}$ is steadily driven upwards by the increasing $F_8$, which reflects the stars' continually decreasing $g$. The increasing $R_{var}$ and $F_8$ values of subgiants and giants on the flicker floor is probably the result of the increasingly important contribution of radial and non-radial pulsations to the overall brightness variations[22,23].

A few stars appear as outliers to the basic picture we have presented here; these are seen towards the right of the vertical cloud of points in our evolutionary diagram (Fig. 3). Some active dwarfs have higher $F_8$ than expected for their $g$ values. Frequent strong flares can boost $F_8$ as currently defined, and some hotter dwarfs are pulsators with enough power near 8 hours to increase their $F_8$ values. A few such cases appear also in the asteroseismic sample (Fig. 1). Some lower $g$ stars have $R_{var}$ above the flicker floor owing to the presence of magnetic activity[24], slow radial pulsations or secular drifts. Finally, a few outliers are simply due to data anomalies. As our technique is refined, these exceptions should be treated carefully before assigning a $F_8$-based $g$ value, particularly for high-$F_8$ stars for which $R_{var}$ is greater than ~3ppt. They constitute a small fraction of the bulk sample, and most of them can be identified as one of the above cases.

Common to all of the stars along the flicker floor is the virtual absence of spot activity as compared to their higher $R_{var}$ counterparts; short-time scale phenomena such as granulation and oscil-lations dominate the brightness variations. Given that spots probably suppress acoustic oscillations in the Sun and other dwarf stars[5,26,27,28], the large $X_0$ of stars along this sequence may partly reflect the ability of short-time scale processes to manifest more strongly now that large spots no longer impede them, along with the increasing complexity of the convective variations. As the stars evolve into full-fledged red giants and beyond, the principal periodicity in their brightness variations increasingly reflects shorter-period oscillations, as opposed to their inherently long-period rotation, because oscillations become dominant over magnetic spots.

It may be possible to differentiate between stars with similar *g* but different internal structures (e.g., first-ascent red giants versus helium burning giants) through application of a sliding timescale of $F_8$ as a function of *g*, where the sliding timescale would capture the changing physical granulation timescales with evolutionary state[20]. Moreover, the behavior of stars on the flicker floor may explain the source of radial velocity "jitter" that now hampers planet detection through radial velocity measurements[28].

**Acknowledgements**

The research described in this paper makes use of Filtergraph[29], an online data visualization tool developed at Vanderbilt University through the Vanderbilt Initiative in Data-intensive Astrophysics (VIDA). We acknowledge discussions with Phillip Cargile, Kenneth Carpenter, William Chaplin, Daniel Huber, Martin Paegert, Manodeep Sinha and David Weintraub. We thank Daniel Huber and Travis Metcalfe for sharing the average asteroseismic parameters of Kepler stars with us. F. A. B. acknowledges support from a NASA Harriet Jenkins Fellowship and a Vanderbilt Provost Graduate Fellowship.

**Author Contributions**

F. A. B. and K. G. S. contributed equally to the identification and analysis of the major correlations. F. A. B. principally wrote the first version of the manuscript. K. G. S. prepared the figures. G. B. calculated the variability statistics of the Kepler light curves and performed an independent check of the analysis. J. P. checked against biases in the datasets. All authors contributed to the interpretation of the results and to the final manuscript.

**Author Information**

Reprints and permissions information is available at www.nature.com/reprints.  The authors declare no competing financial interests.  Correspondence and requests for materials should be addressed to F. A. B. at fabienne.a.bastien@vanderbilt.edu

**Supplementary Information** is linked to the online version of the paper at www.nature.com/nature.

**Figures**

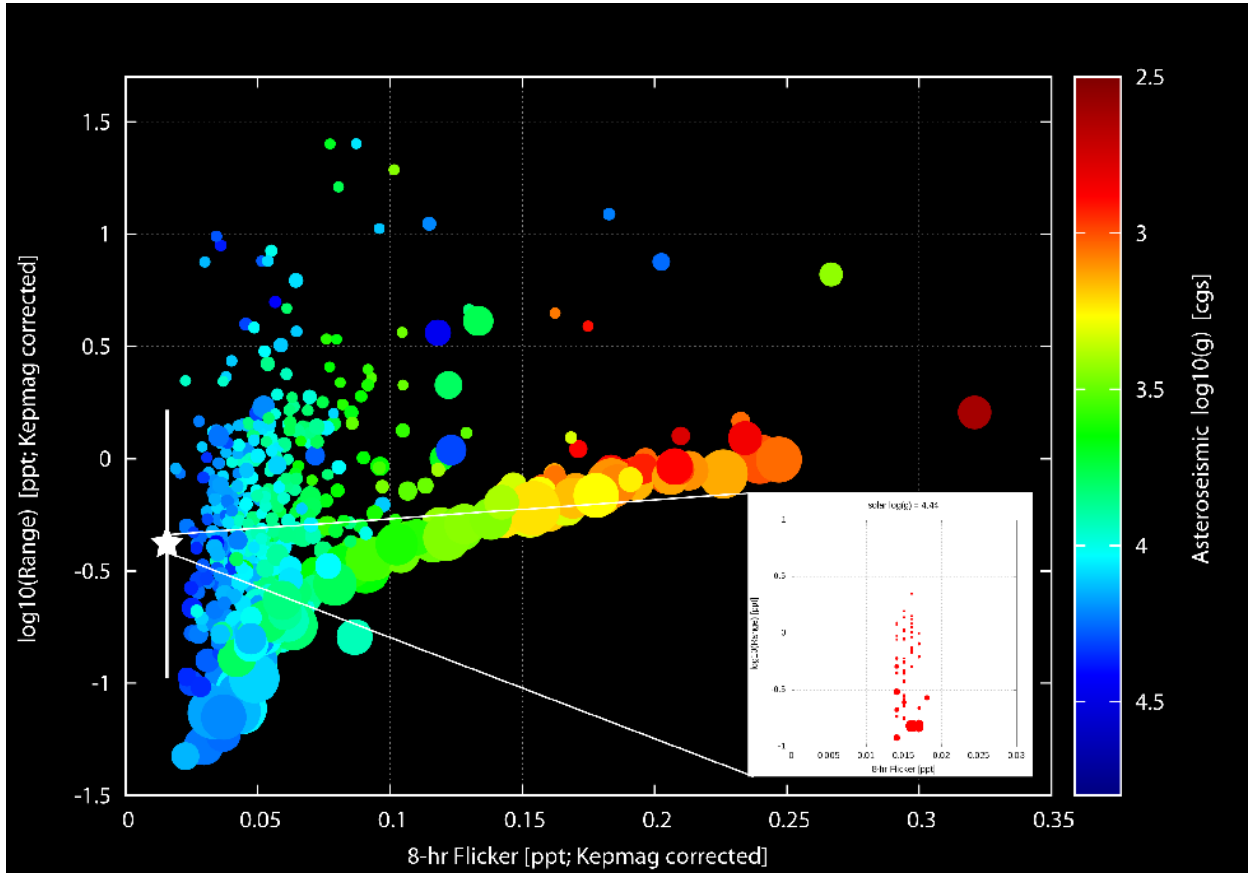

Figure 1: Simple measures of brightness variations reveal a fundamental "flicker sequence" of stellar evolution. We establish the evolutionary states of stars with three simple measures of brightness variations[8]. The abscissa, 8-hr flicker ($F_8$), measures brightness variations on time scales of 8 hours or less. The ordinate, $R_{var}$, yields the largest amplitude of the photometric variations in a 90-day timeframe. $X_0$ (symbol size; ranging from 0.01 to 2.1 crossings per day), conveys the large-scale complexity of the light curve. We correct both $R_{var}$ and $F_8$ for their dependence on Kepler magnitude ("Kepmag"). Color represents asteroseismically determined $g$. We observe two populations of stars: a vertical cloud composed of high-$g$ dwarfs and some subgiants, and a tight sequence, the flicker floor, spanning an extent in $g$ from dwarfs to giants. The typically large $R_{var}$ values of stars in the cloud, coupled with their simpler light curves (small $X_0$), implies brightness variations driven by rotational modulation of spots. In contrast, large $X_0$ values characterize stars on the sequence. The $F_8$ values of stars in this sequence increase inversely with $g$ because the physical source of $F_8$ is sensitive to $g$. $R_{var}$ also increases with $F_8$ along the floor, because $F_8$ is a primary contributor to $R_{var}$ (as opposed to starspots above the floor). Stars with a given $F_8$ value cannot have $R_{var}$ less than that implied by $F_8$ itself: quiet stars accumulate on the flicker floor because they are prevented from going below it by the statistical definition of the two quantities. Stars above the floor have larger amplitude variations on longer time scales that set $R_{var}$. The large star symbol with vertical bars and the inset show the Sun's behavior over the course of its 11-year magnetic cycle. The Sun's $F_8$ value is largely invariant over the course of its cycle, just as its $g$ value is invariant.

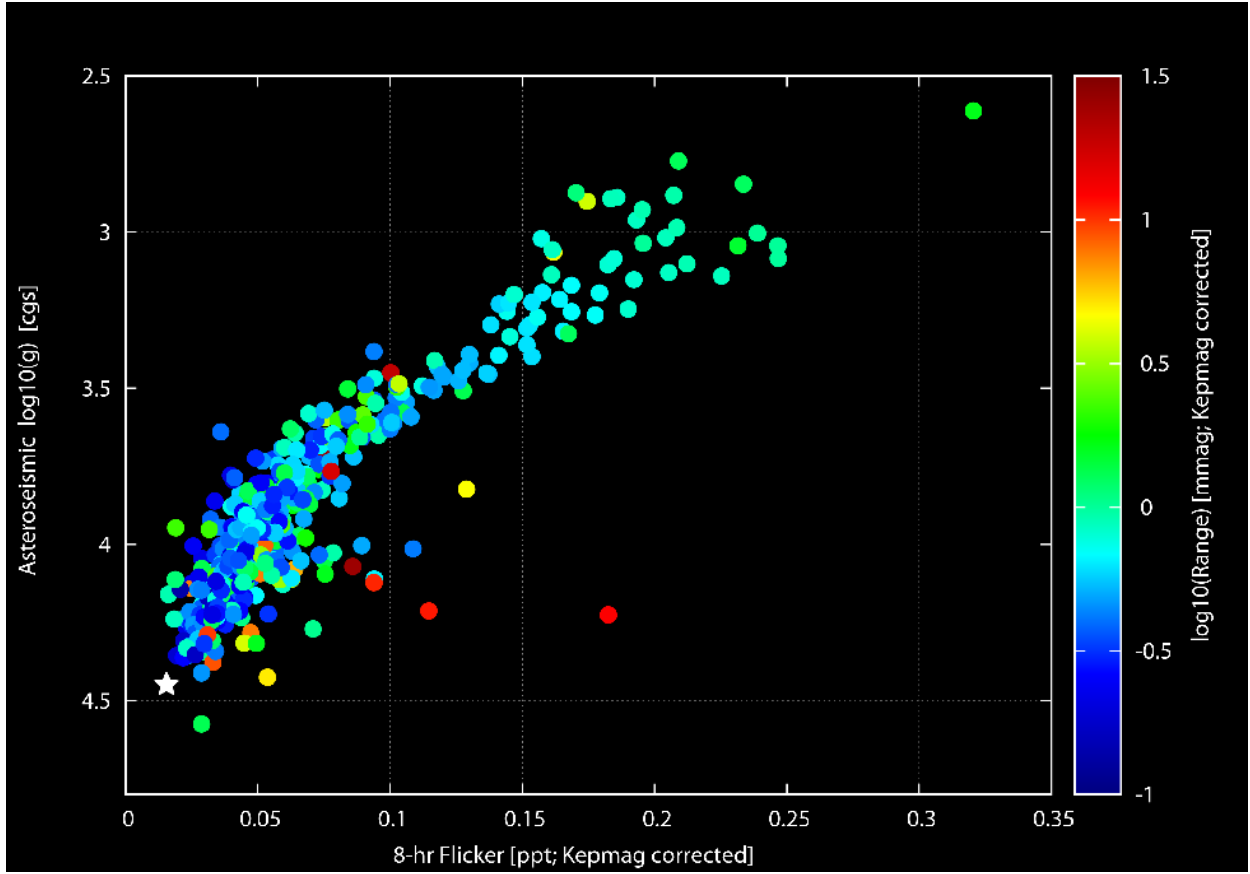

Figure 2: Stellar surface gravity manifests in a simple measure of brightness variations. The same stars from Fig. 1 with Kepler Quarter 9 data. Asteroseismically determined[4] $g$ shows a tight correlation with $F_8$. Color represents the $R_{var}$ of the stars' brightness variations; outliers tend to have large brightness variations. Excluding these outliers, a cubic polynomial fit through the Kepler stars and through the Sun (large star symbol) shows a median absolute deviation of 0.06 dex and a root-mean-square deviation of 0.10 dex (Supplementary Information). To simulate how the solar $g$ would appear in the archival data we use to measure $g$ for other stars, we divide the solar data into 90-day "quarters." Our $F_8$-$g$ relation measured over multiple quarters then yields a median solar $g$ of 4.442 with a median absolute deviation of 0.005 dex and an RMS error of 0.009 dex (the true solar $g$ is 4.438).

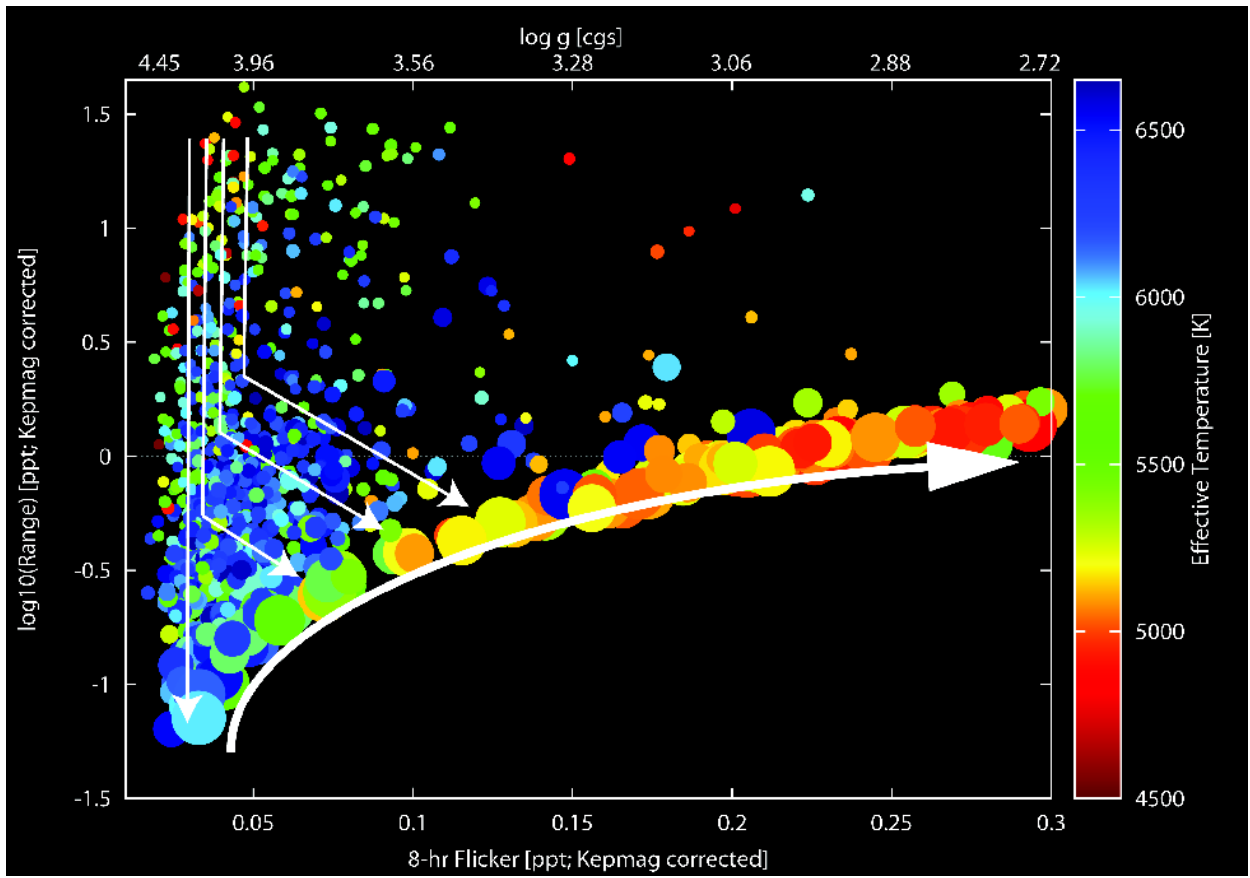

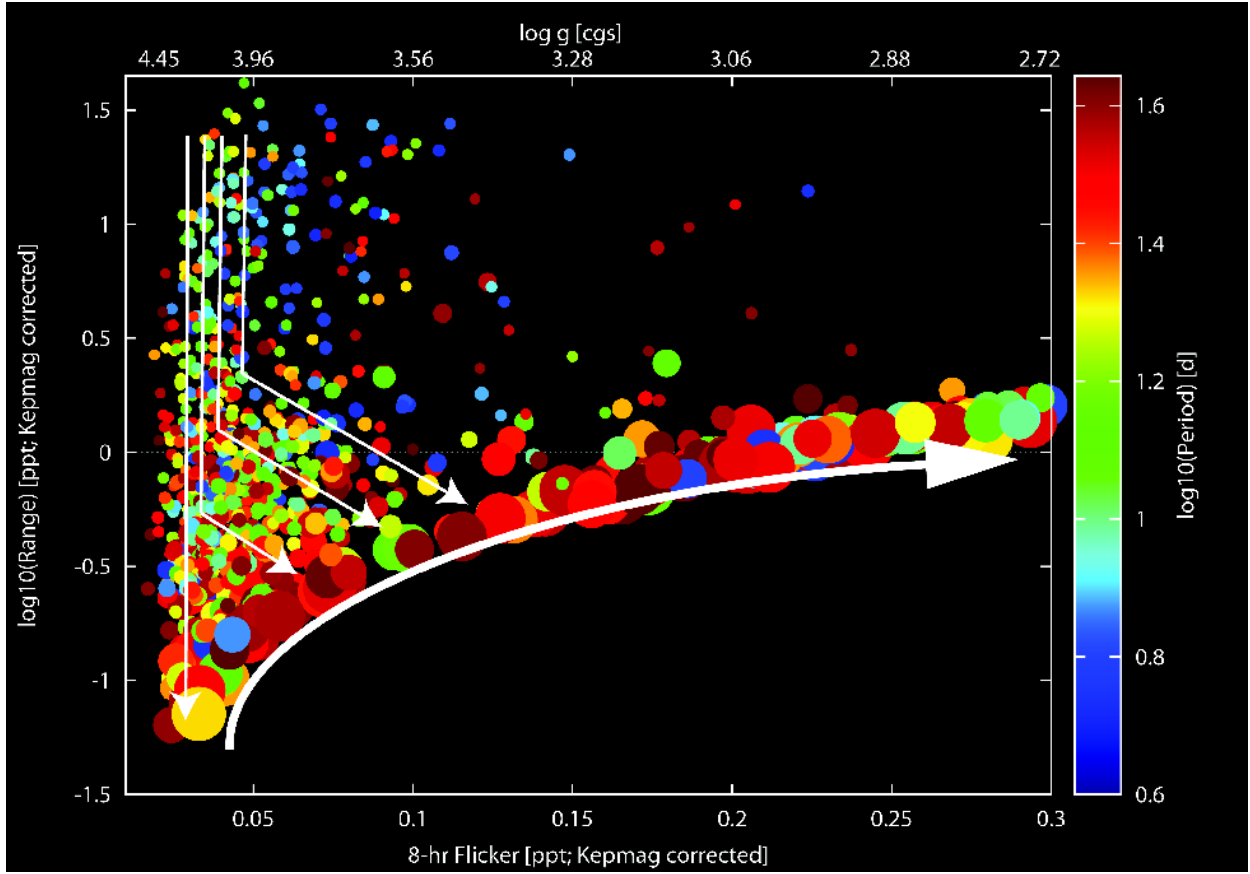

Figure 3: An integrative view of stellar evolution in a new diagram of brightness variations. Same as Figure 1 but for Kepler stars lacking asteroseismic $g$. We include a $g$ scale at the top (from conversion of the $F_8$ scale at bottom via our calibrated relationship). Here, we selected stars with Kepler magnitudes between 11.0 and 11.85 in order to limit the sample to ~1000 stars for visual clarity (1,012 points are shown). We removed objects that are potentially blended (Kepler flux contamination greater than 0.05) as well as those that may be galaxies (Kepler star-galaxy flag other than 0). Arrows schematically indicate the evolutionary paths of Sun-like stars in this diagram. Stars generally move from top to bottom, as the overall brightness fluctuations due to spots decrease with time, and then from left to right as their $g$ values decrease. All stars even-

tually arrive on the flicker floor sequence and evolve along it. *Top:* Color represents effective temperature. Stars cool as they evolve from left to right, from dwarfs to red giants. We restricted the effective temperatures to be 4500-6650 K, using the revised temperature scale for Kepler stars[30]. *Bottom:* Same as top, but color-coded by the dominant periodicity in the light curve. We limited the sample to stars with dominant periods longer than 3 days (to eliminate very rapidly rotating active stars) and shorter than 45 days (half the Kepler 90-day data interval). This period traces rotation for unevolved stars and pulsations for evolved ones. Dwarfs generally show the expected spin-down sequence with decreasing $R_{var}$ (correlated with the level of surface magnetic activity). Subgiants and giants broadly display very slow rotation as expected.

**Supplementary Information**

Asteroseismic measurement of surface gravities (*g*):

We used as our sample the first ensemble asteroseismic analysis of Kepler stars[4]. The $\Delta\nu$ and $\nu_{max}$ values from that work were shared with us privately by the authors (D. Huber and T. Metcalfe, priv. comm.). $\nu_{max}$ refers to the central frequency of the solar-like p-mode oscillations, where the power is greatest, while $\Delta\nu$, the large frequency separation, measures the sound travel time across the star's diameter[20]. Taking these, together with the revised stellar effective temperatures[30], we applied the standard scaling relations that transform them into mass and radius[4] to calculate *g*. Seismic parameters are now available for several thousand stars[5,6] and will be used in a future calibration of our $F_8$-*g* relation.

Detailed description of how each of the photometric variability measures is calculated

We employ the following three variability statistics[8] in this work. We use as our starting point for all of these Kepler Quarter 9 PDC-MAP data:

- Range ($R_{var}$): obtained by sorting the pipeline-reduced Kepler light curve by differential intensity and measuring the range between the 5th and 95th percentile. This quantity traces the stellar surface spot coverage.
- Number of zero crossings ($X_0$): computed by smoothing the light curve by 10 hour (20 point) bins and counting the number of times the resultant light curve crosses its median value. It provides an assessment of the complexity of the light curve. For example, spots produce variations larger than the high-frequency noise resulting in a small $X_0$
- 8-hour flicker ($F_8$): determined by performing a 16-point (8 hour) boxcar smoothing of the light curve, subtracting it from the original light curve and measuring the root-mean-square (RMS) of the result. We handle data gaps by interpolating across any missing 30-minute data bins. The result is a somewhat decreased $F_8$ because it is zero in these interpolated segments. In practice the data gaps are so few and small that the impact on the overall $F_8$ is negligible. We, at present, do not employ sigma-clipping. There are rare cases in which a few light curve points are extreme enough to boost the RMS, but are clipped from the $R_{var}$. Such cases can appear below the "flicker floor" we have described. Examples include a quiet star with a deep transiting planet, or one very large instrumental spike affecting only a few points. $F_8$ measures stellar variability on time scales of 8 hours or less.

Details on how the solar data were put into "Kepler equivalent" form

We used SOHO Virgo[16] light curves, whose passband is similar to that of Kepler[17]. We took light curves spanning an entire solar cycle to examine the influence of changing spot activity (i.e. changing $R_{var}$) over the course of a stellar magnetic cycle on our findings. We divided the SOHO light curve into 90 day segments, to simulate the length of a Kepler "quarter," and we sampled each segment to achieve an effective cadence of 30 minutes, similar to the cadence of the Kepler long-cadence light curves.

We note that the actual derived $F_8$ depends on the filter used and the treatment of the solar data. The solar brightness variations are largest in the blue filter, moderate in the green filter, and smallest in the red filter (as expected for the temperature of the Sun). One finds, for instance, a roughly 30% larger $F_8$ than the value reported here when considering solely the green filter data.

Thus, previous analyses have used a sum of the green and red filter data (which is more nearly the same as a broadband filter), and we now use the TSI "white" data (which is also a broadband realization of the solar variations).

Figure S1: Kepler Magnitude correction:

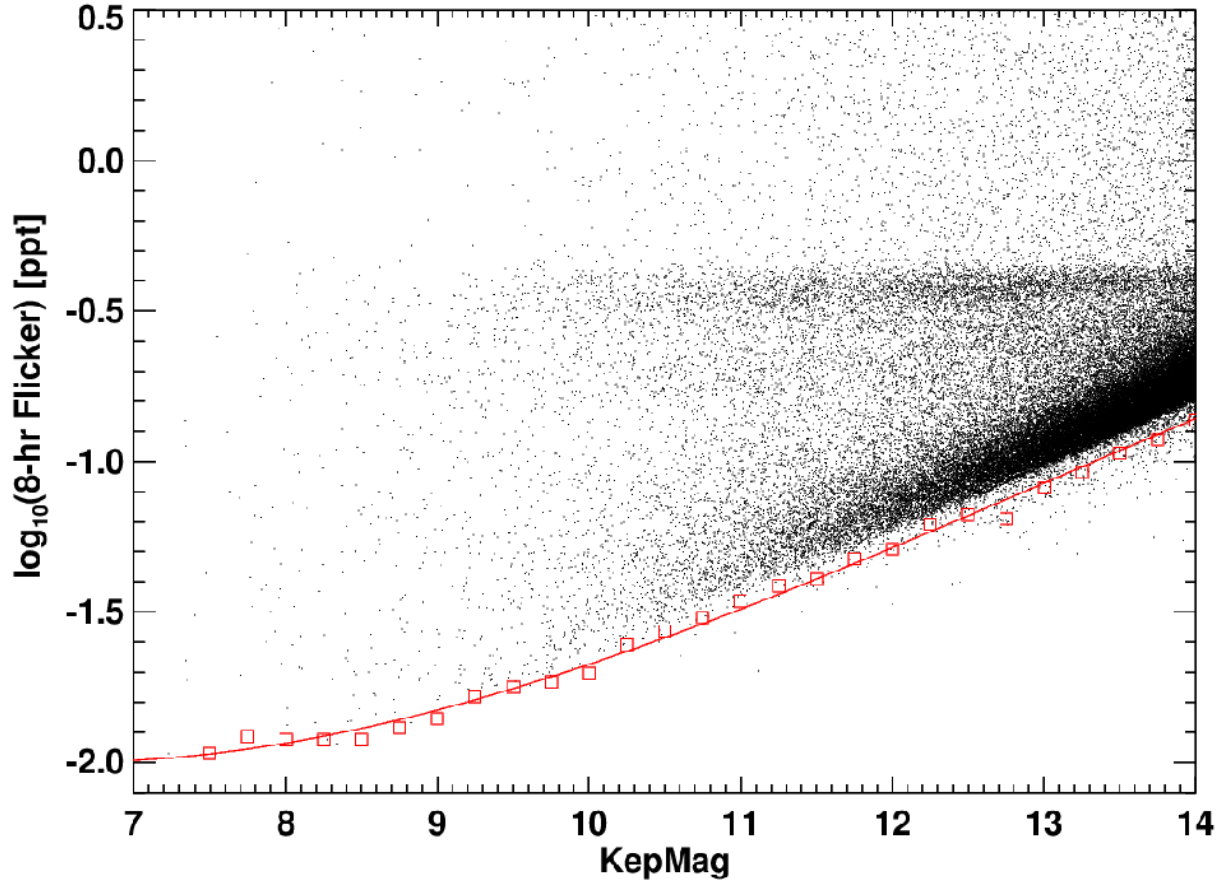

The measures of brightness variations that we use all show dependencies on stellar brightness. Fainter stars naturally exhibit larger brightness variability in $R_{var}$ and $F_8$ simply due to increased photon noise. Not accounting for this results in an overestimate of $F_8$ (equivalently, an underestimate of $g$). We therefore correct these measures using empirical relations obtained from the entire sample of Kepler Quarter 9 light curves. We fit each of the brightness variability measures versus stellar apparent magnitude in the Kepler bandpass ("Kepmag" or $K_p$) using a simple 4th order polynomial fit to the lower envelope of points, defined as the bottom 0.5-th percentile of points in 0.1 magnitude wide bins. These polynomial relations were then subtracted in quadrature from the measured brightness variation measures, and these corrected variability measures are identified in all figures as "Kepmag corrected." The final Kepler magnitude relation used in this work is:

$$\min\left(\log_{10} F_8\right) = -0.03910 - 0.67187\, K_p + 0.06839\, K_p^2 - 0.001755\, K_p^3$$

where $K_p$ is the Kepler magnitude, and the fit applies for $7 < K_p < 14$. The final $F_8$ that we use is obtained by subtracting this $\min(F_8)$ from the measured $F_8$ in quadrature (the quadrature subtraction is performed linearly, not logarithmically).

Figure S2: Details on $g$ versus $F_8$ fit relation:

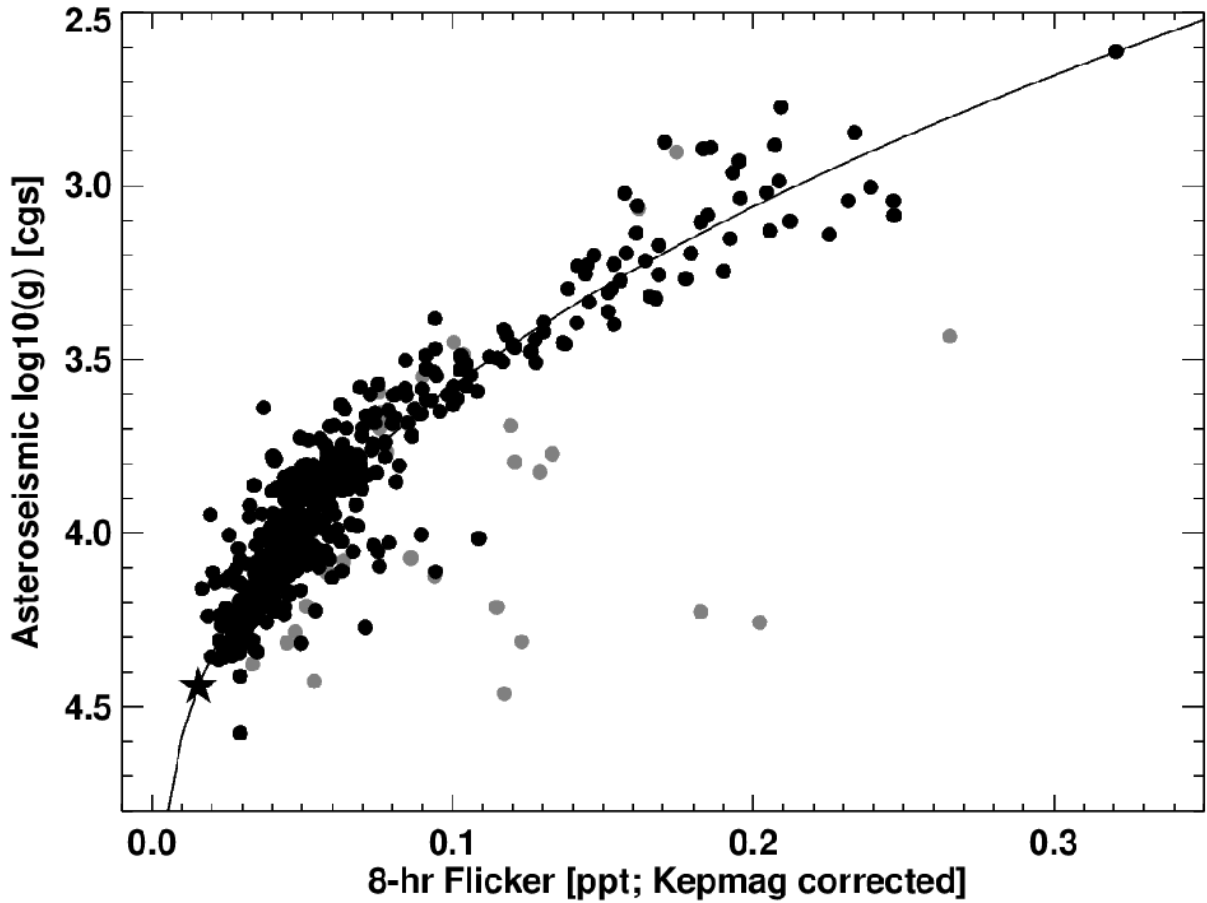

We draw our "gold standard" sample from the first published asteroseismic analysis of Kepler stars[5] (we note that there are now over 10,000 seismically analyzed Kepler stars that will permit our relations to be extended further in future work). From these, we used as our base sample the 542 stars (gray points) possessing both asteroseismically determined masses and radii and Quarter 9 long-cadence light curves from which we could compute our variability statistics[8]. These stars have Kepler magnitudes brighter than 12, effective temperatures[30] between 4500 < Teff < 6650 K, and a Kepler flux contamination flag of less than 0.05. Most of the outliers from the trend are stars with very short periods (likely short-period pulsators or rapidly rotating active stars) and/or large brightness excursions ($R_{var}$). Therefore, we removed from the polynomial fit 10 stars with Period < 3 days and 19 additional stars with $R_{var}$ > 2.5 ppt. The remaining 503 stars (black) were fitted with a cubic polynomial (solid curve). $F_8$ was corrected for the dependence on Kepler magnitude as above. The large star symbol at lower left represents the Sun with $g$ = 4.44 and a median $F_8$ of 0.015 ppt over the entire 11-year solar activity cycle. The polynomial fit was forced through the solar value since there are few asteroseismic stars with $g$ as high as the Sun, however the fit passes within 0.05 dex of the solar value even without forcing the fit. The final polynomial fit relation is:

$$\log_{10} g = 1.15136 - 3.59637\, x - 1.40002\, x^2 - 0.22993\, x^3$$

where $x = \log_{10}(F_8)$ and $F_8$ is in units of ppt. The root-mean-square of the $g$ residuals about the polynomial fit is 0.10 dex and the median absolute deviation is 0.06 dex.

Figure S3: Examples of light curves in different regions of the photometric variability evolutionary diagram:
*Top:* The photometric evolutionary diagram, showing the regions from which we draw our sample light curves. *Middle:* Six examples of light curves from Quarter 9 with different $R_{var}$ and $F_8$. The black curves show the differential intensity in units of parts per thousand. The red curves show the result of applying a boxcar smoothing of 16 points (8 hours). Thus, the $F_8$ is the RMS difference between the black and red curves. The first three stars are taken from the left hand edge of Figure S3, *top*, at the top, middle, and bottom of the dwarf stars (labeled 1-3). Though most stars in this region are likely dwarfs, a small fraction of giants with very low $g$ (typically also with very large $R_{var}$) contaminate this region. One could enhance the determination of $g$ for dwarfs with $R_{var} > 2.5$ ppt by considering additional diagnostics present in the same light curve data. For example, the true dwarfs that dominate the upper left part of the diagram have large $R_{var}$ and are therefore active. They thus should exhibit strong periodicity on timescales expected for rotation of active dwarfs (i.e., Prot < 20 d). The second set of three stars are at evenly spaced locations along the flicker floor in that figure, moving out to higher $F_8$ and $R_{var}$ (labeled 4-6). The Kepler IDs of each star are indicated. *Bottom:* A 10-day section of the light curves, to bring out more details. The $R_{var}$ and $F_8$ values for the six stars are listed here also for reference (cf. Table below). We also include the temperatures[30] and the $g$ from both the Kepler Input Catalog and our $F_8$ relation.

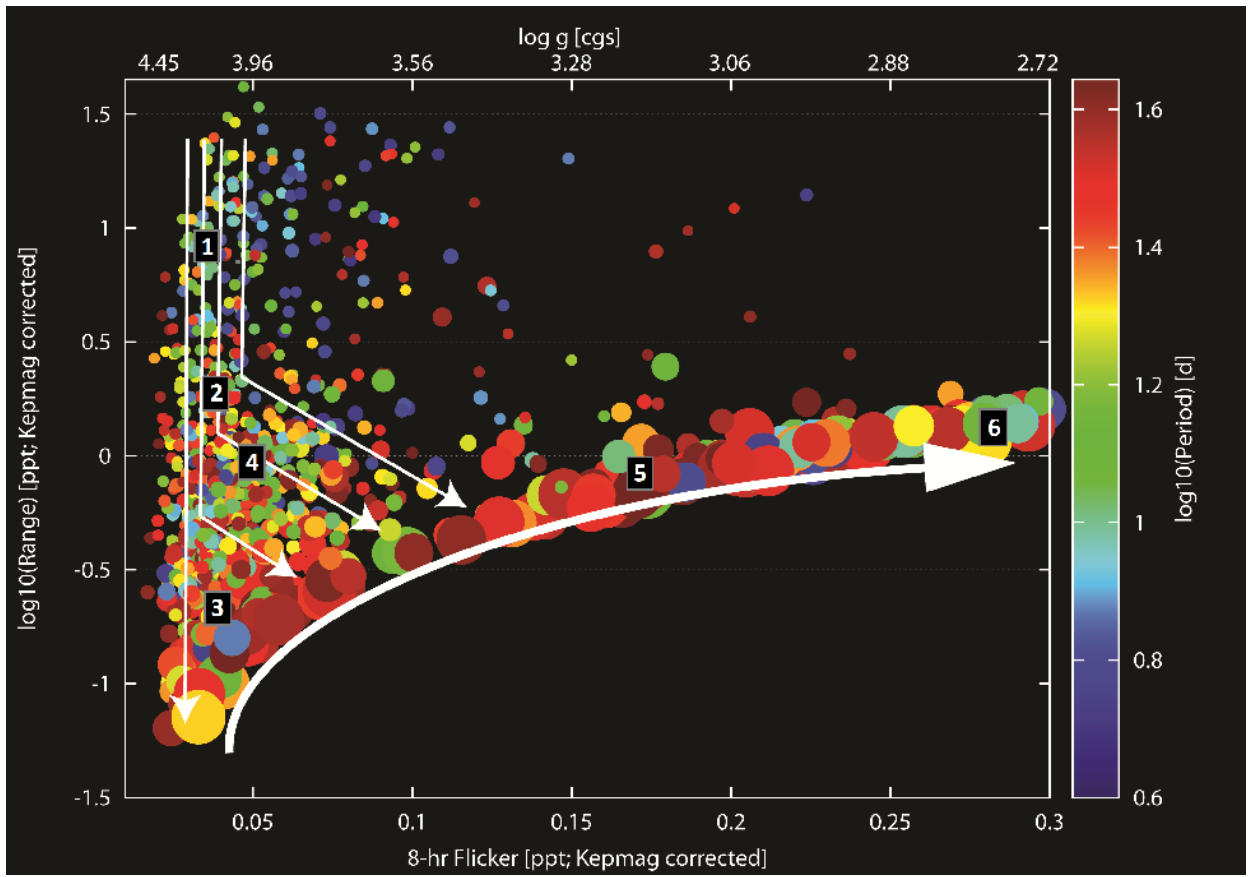

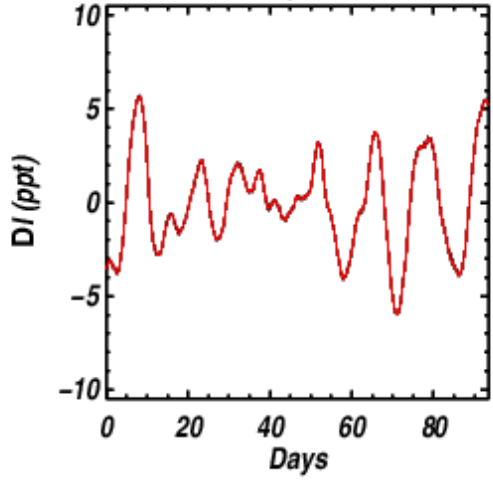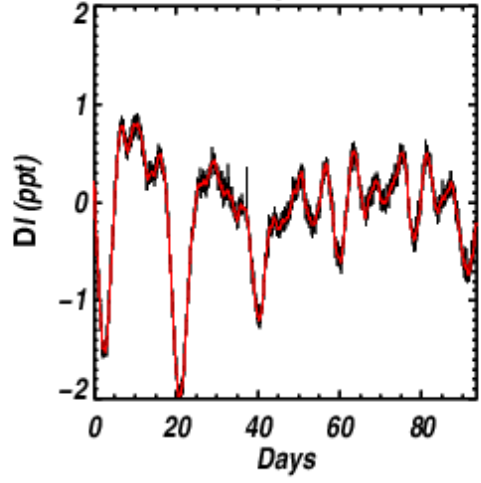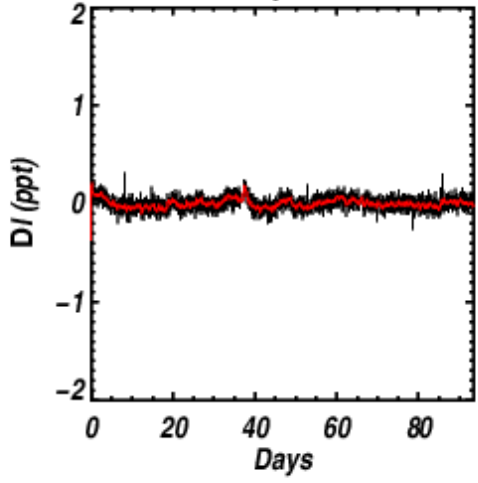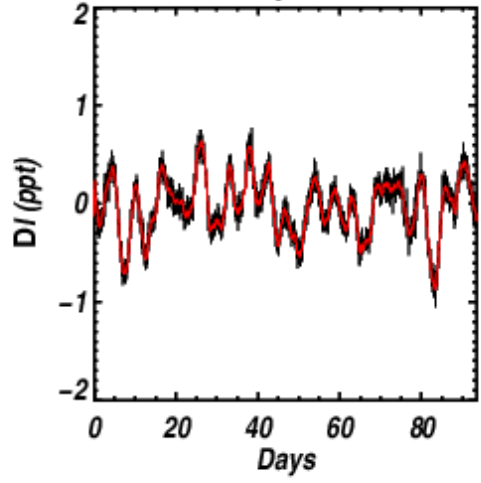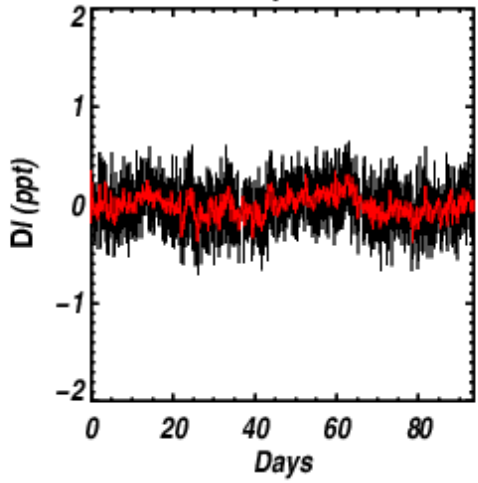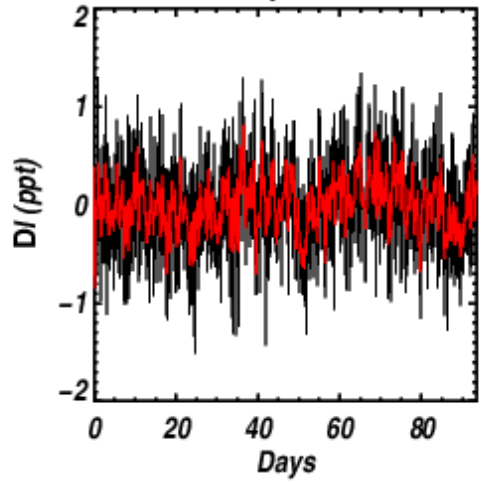

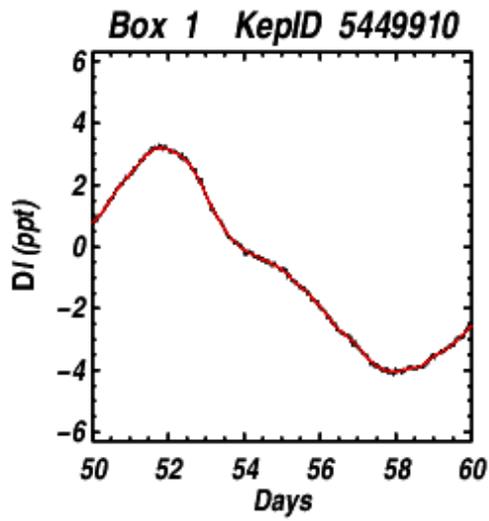
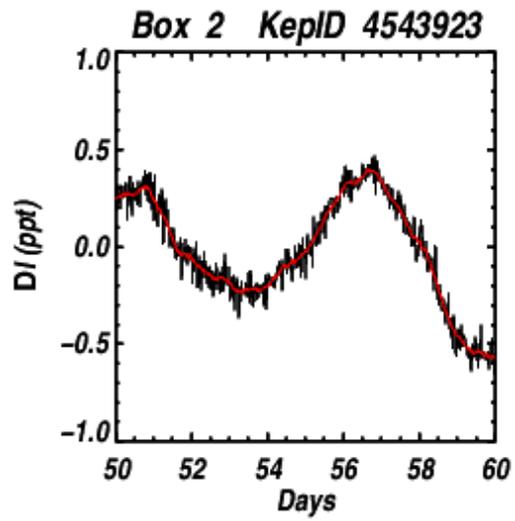
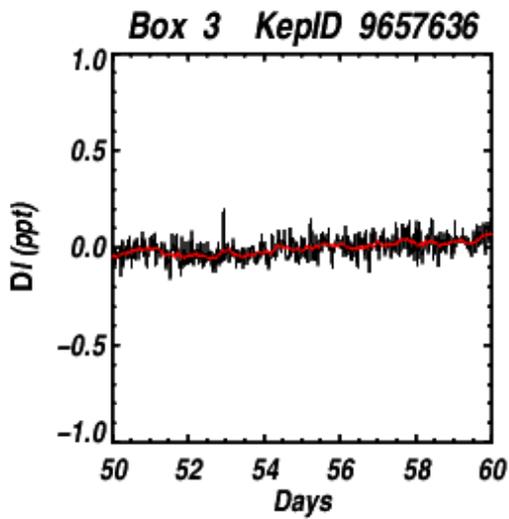
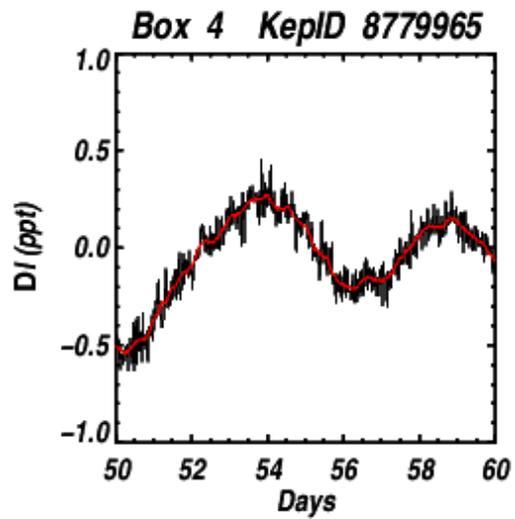
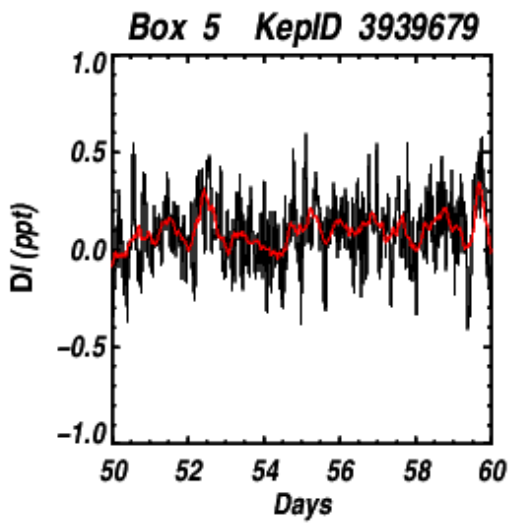
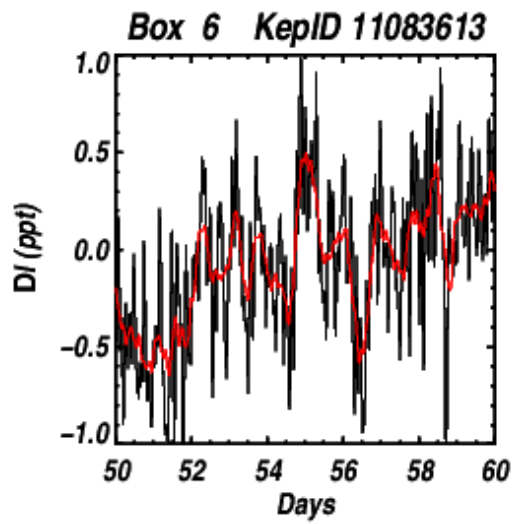

| Star | 1 | 2 | 3 | 4 | 5 | 6 |
|---|---|---|---|---|---|---|
| Kepler ID | 5449910 | 4543923 | 9657636 | 8779965 | 3939679 | 11083613 |
| Effective Temperature[30] (K) | 5417 | 6062 | 6153 | 5799 | 5036 | 5062 |
| Kepler Magnitude | 11.235 | 11.623 | 11.253 | 11.730 | 11.099 | 11.629 |
| $R_{var}$ (ppt) | 8.561 | 1.841 | 0.206 | 0.956 | 0.707 | 1.359 |
| $F_8$ (ppt) | 0.046 | 0.055 | 0.049 | 0.067 | 0.171 | 0.284 |
| Magnitude-corrected $F_8$ (ppt) | 0.029 | 0.034 | 0.033 | 0.049 | 0.171 | 0.284 |
| $g$ (cgs) from KIC | 4.539 | 4.248 | 4.209 | 3.578 | 4.248 | 3.361 |
| $g$ (cgs) from $F_8$ | 4.212 | 4.141 | 4.155 | 3.974 | 3.191 | 2.738 |